\pdfoutput=1
\RequirePackage{ifpdf}
\ifpdf 
\documentclass[pdftex]{sigma}
\else
\documentclass{sigma}
\fi

\newcommand{\call}{{\mathcal L}}

\newcommand{\CC}{{\mathbb C}}
\newcommand{\QQ}{{\mathbb Q}}
\newcommand{\ZZ}{{\mathbb Z}}
\newcommand{\PP}{{\mathbb P}}

\begin{document}

\allowdisplaybreaks

\renewcommand{\thefootnote}{$\star$}

\renewcommand{\PaperNumber}{056}

\FirstPageHeading

\ShortArticleName{Monodromy of an Inhomogeneous Picard--Fuchs Equation}

\ArticleName{Monodromy of an Inhomogeneous\\ Picard--Fuchs Equation\footnote{This
paper is a contribution to the Special Issue ``Mirror Symmetry and Related Topics''. The full collection is available at \href{http://www.emis.de/journals/SIGMA/mirror_symmetry.html}{http://www.emis.de/journals/SIGMA/mirror\_{}symmetry.html}}}

\Author{Guillaume LAPORTE~$^\dag$ and Johannes WALCHER~$^{\dag\ddag}$}

\AuthorNameForHeading{G.~Laporte and J.~Walcher}

\Address{$^\dag$~Department of Physics, McGill University, Montr\'eal,
Qu\'ebec, Canada}
\EmailD{\href{mailto:guillaume.laporte@mail.mcgill.ca}{guillaume.laporte@mail.mcgill.ca}}

\Address{$^\ddag$~Department of Mathematics and Statistics, McGill University,
Montr\'eal, Qu\'ebec, Canada}
\EmailD{\href{mailto:johannes.walcher@mcgill.ca}{johannes.walcher@mcgill.ca}}

\ArticleDates{Received June 08, 2012, in f\/inal form August 20, 2012; Published online August 22, 2012}

\Abstract{The global behaviour of the normal function associated with van Geemen's
family of lines on the mirror quintic is studied. Based on the associated
inhomogeneous Picard--Fuchs equation, the series expansions around large complex
structure, conifold, and around the open string discriminant
are obtained. The monodromies are explicitly calculated from this data and checked
to be integral. The limiting value of the normal function at large complex structure
is an irrational number expressible in terms of the di-logarithm.}

\Keywords{algebraic cycles; mirror symmetry; quintic threefold}

\Classification{14C25; 14J33}

\vspace{-3mm}

\section{Introduction}

The early mathematical literature on mirror symmetry is replete with attempts to
elucidate the enumerative predictions made by physics, \cite{cdgp} about rational curves
on Calabi--Yau threefolds by utilizing the understanding of algebraic cycles on
higher-dimensional varieties obtained from transcendental methods and deformation theory.
The development of Gromov--Witten theory in the mid 1990's, culminating in a verif\/ication
of the predictions, clarif\/ied the separation of the enumerative aspects of mirror symmetry
(A-model) from the Hodge theoretic ones (B-model). Later, with the emergence of homological
mirror symmetry (HMS) and the Strominger--Yau--Zaslow (SYZ) picture, the classical questions
had been all but driven out, and the enumerative aspects relegated to some challenging
combinatorics of toric manifolds.

In recent years, it has become clear that algebraic cycles in fact have an eminent role
to play in relating the established theory underlying classical mirror
symmetry~-- Gromov--Witten theory and variation of Hodge structure, with the more
elaborate versions such as HMS and SYZ. Indeed, if computing and comparing geometric
invariants is the primary goal in elucidating the physics associated with Calabi--Yau
manifolds, it is rather natural that algebraic cycles should be considered soon after
the crudest topological data. They are obvious and well-studied algebraic invariants
of the derived category, and have been proposed as mirrors to the (far less obvious, in fact
not yet def\/ined) on-shell invariants of the Fukaya category.

On the other hand, when viewed against the backdrop of the intricacies of the string
duality web, and the fact that some of the deepest conjectures in algebraic geometry
concern algebraic cycles, the speculation that the entire enumerative geometry of
Calabi--Yau manifolds (Gromov--Witten, BPS, and otherwise) might be encoded in the
(higher) Chow groups of the mirror manifold seems rather stably grounded. The earliest
reference to such ideas that we are aware of is made in~\cite{doma}.

A most intriguing consequence of this basic assumption arises via the fundamentally
arithmetic nature of algebraic cycles. Even if (as would be natural for a physicist,
interested in the real world or not), one is expecting to deal primarily with complex
numbers, the f\/ield of def\/inition of the algebraic cycle relative to that of the underlying
variety invariably enters the discussion in all but the very simplest situations.
Whether this is a mere curiosity in a rather special setup, or the hint of a
more signif\/icant connection between the two subjects,
it is clear that some adjustement of the physical picture
will have to take place. And quite
similarly, symplectic geometers should need to contemplate arithmetic considerations playing
an important role in a detailed study of certain Fukaya categories and their deformations.

The exploration of these questions was taken up in the recent paper~\cite{arithmetics},
using explicitly constructed algebraic cycles on the mirror quintic threefold.
Without identifying the precise A-model setup, it was found that the would-be
enumerative invariants are in general algebraic numbers satisfying a certain
integrality condition. Actually, the (irrational) algebraicity of the relevant
expansion coef\/f\/icients is an inevitable consequence of the fact that the irreducible
components of the algebraic cycle are def\/ined over an extension of the f\/ield of
def\/inition of the underlying manifold (in this case, the f\/ield $\QQ$ of rational
numbers). The meaning of the integra\-li\-ty and the A-model or spacetime physics
explanation for the irrationality of the enumerative invariants, remains to be found.

A dif\/ferent type of arithmeticity of the same Hodge theoretic invariants of algebraic cycles
has been recently explained by Grif\/f\/iths, Green and Kerr, see~\cite{ggk}.
In certain degenerate
limits, including those of relevance for mirror symmetry, standard Abel--Jacobi mappings
such as those underlying the calculations of the D-brane superpotential carried out in~\cite{arithmetics} turn out to be given by higher Abel--Jacobi mappings
on certain higher Chow groups. In turn, these degenerate Abel--Jacobi mappings are regulators on
certain K-groups of the singular member of the family, and have an arithmetic signif\/icance.
This was illustrated in~\cite{ggk} in various examples, including one closely
related to the one we study here.

In this paper, we will report on some further calculations around one of the algebraic
cycles on the mirror quintic studied in~\cite{arithmetics}. Starting from the
inhomogeneous Picard--Fuchs equation satisf\/ied by the associated normal function, we will
compute the expansion of that normal function around the various interesting points
in the complex structure moduli space. By comparing (numerically) these various
expansions, we will be able to determine the complete analytic continuation of the normal
function around the entire moduli space. In particular, this will provide the
transformation under mondromy around the singular loci. The monodromies will be
integral, as they should be on general grounds. This can be viewed as a consistency
check on the normalization of the normal function found in~\cite{arithmetics}.
(This check is satisfying, but honestly redundant, as the normalization is completely
determined by the methods used in~\cite{arithmetics}.)

Our calculations will also determine a constant of integration that was not calculated
in~\cite{arithmetics}. This interesting constant is the only one not constrained
to be rational by monodromy conside\-ra\-tions. It is the limiting value of the normal
function, and precisely the one number which the invariants of~\cite{ggk} reduce
to for co-dimension 2 cycles on Calabi--Yau threefolds. Conf\/irming general expectations,
and results of~\cite{ggk}, we will f\/ind that this constant is, up to a rational
multiple, given by the special value of an L-function of the algebraic number f\/ield
over which the underlying cycle is def\/ined.

\section{Equations}

\begin{figure}[t]
\centering
\includegraphics{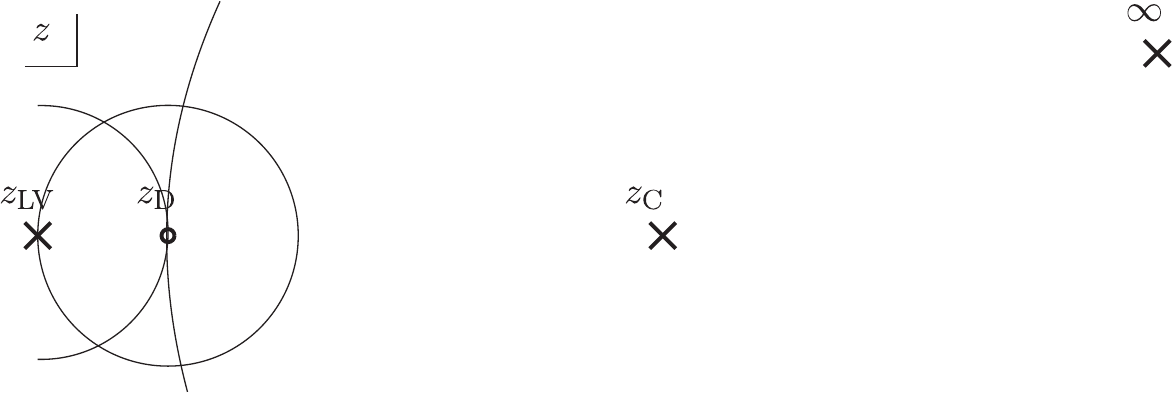}
\caption{Moduli space of the mirror quintic with van Geemen family. The arcs indicate
the circles of convergence of the series expansions about the various singular points.}
\label{space}
\end{figure}

We start from the vanishing locus $\{W=0\}$ of the familiar family of polynomials
\begin{gather*}
W= x_1^5+x_2^5+x_3^5+x_4^5+x_5^5 - 5\psi x_1x_2x_3x_4x_5
\end{gather*}
in $\PP^4\ni[x_1:x_2:x_3:x_4:x_5]$. The mirror quintic family is the resolution of the quotient
\begin{gather*}
Y = \{W = 0 \}/(\mu_5)^3,
\end{gather*}
where $(\mu_5)^3\cong (\ZZ/5\ZZ)^3$ is the Greene--Plesser group of phase symmetries
leaving $W$ invariant. As~$\psi$ varies, it parameterizes local deformations of the
complex structure of the mirror quintic. Multiplication of $\psi$ by a f\/ifth root
of unity can be undone by a change of coordinates on~${\mathbb P}^4$, and a good global
coordinate on the moduli space is the familiar
\begin{gather*}
z = (5\psi)^{-5}.
\end{gather*}
Our moduli space has three distinguished points: the large volume point, corresponding
to $z=z_{\rm LV}=0$, the conifold point at $z=z_{\rm C}=5^{-5}$, and the orbifold (or Gepner)
point, $z^{-1}={z_{\rm G}}^{-1}=0$, see Fig.~\ref{space} for a sketch.

The convenient choice of holomorphic three-form on $Y$ comes from
\begin{gather*}
\Omega = \left(\frac{5}{2\pi i}\right)^3\mathop{\operatorname{Res}}\limits_{W=0}
\frac{\sum\limits_{i=1}^5 (-1)^i x_i dx_1\wedge\cdots
\wedge\widehat{dx_i}\wedge \cdots dx_5}{W}.
\end{gather*}
Then, for any $[\Gamma]\in H_3(Y,\ZZ)\cong\ZZ^4$, the corresponding period
\begin{gather*}
\varpi(z) = \int_\Gamma\Omega
\end{gather*}
satisf\/ies as a function of $z$ the dif\/ferential equation
\begin{gather}
\label{deq}
\call\varpi(z) = 0,
\end{gather}
where $\call$ is the Picard--Fuchs dif\/ferential operator
\begin{gather}
 \label{oPF}
\call = \theta^4 - 5z(5\theta+1)(5\theta+2)(5\theta+3)(5\theta+4) ,
\qquad\theta \equiv \frac{z d}{dz}.
\end{gather}
Now let $\omega$ be a non-trivial third root of unity, i.e., a solution of the quadratic
equation
\begin{gather}
\label{third}
1+\omega+\omega^2 = 0.
\end{gather}
Then, as one may check, the holomorphic line in $\PP^4$ given by the equations
\begin{gather}
\label{lines}
x_1+\omega x_2+\omega^2 x_3  =0,\qquad
a(x_1+x_2+x_3)  = 3x_4 , \qquad
b(x_1+x_2+x_3)  = 3x_5
\end{gather}
is contained in the hypersurface $\{W=0\}$, if and only if, for any $\psi$, $a$ and $b$
form a solution to
the equations
\begin{gather}
\label{pars}
ab\psi = 6  ,\qquad a^5+b^5 = 27.
\end{gather}
The lines given by equations~(\ref{lines}) subject to equations~(\ref{pars}), and their images
under the group of discrete symmetries of $W$ are known as van Geemen lines and have
been fruitfully studied over the years, see in particular~\cite{albanokatz,mustatathesis}. One of the results that is relevant for us is that
a~van~Geemen line at f\/ixed $\psi$ belongs to one of two families of lines that can be
distinguished by exchanging $a$ and $b$ or, equivalently, replacing $\omega$ with $\omega^2$ in
equation~(\ref{third}).

Passing the orbit of a van Geemen line under the Greene--Plesser group $(\mu_5)^3$ through the
resolution of singularities induces a family of algebraic cycles on the mirror quintic $Y$
that we shall denote by~$C_\omega$. The truncated normal function associated with this family of
cycles is the integral
\begin{gather}
\label{integral}
\tau(z) = \int_{C_\omega}^{C_{\omega^2}} \Omega
\end{gather}
of the holomorphic three-form against a three-chain bounding $C_{\omega^2}-C_\omega$.
This integral was shown in~\cite{arithmetics} to satisfy the inhomogeneous
Picard--Fuchs equation
\begin{gather}
\label{iPF}
\call\tau(z) = \frac{1+2\omega}{(2\pi i)^2} \cdot \frac{32}{45}
\cdot \frac{\frac{63}{\psi^5} + \frac{1824}{\psi^{10}} -\frac{512}{\psi^{15}}}
{\bigl(1-\frac{128}{3\psi^5}\bigr)^{5/2}}.
\end{gather}
The characteristic denominator originates from the discriminant of the equations~(\ref{pars}). This discriminant consists of the Gepner point, $\psi=0$, as well as
the locus $z=z_{\rm D}=3\cdot 2^{-7}\cdot 5^{-5}$. We will sometimes refer to the
latter as the ``open string discriminant'', keeping in mind that the Gepner point
is the other branch point of the quadratic extension of the moduli space by the
inhomogeneity. Note in particular that while at f\/ixed $\psi$ or $z$,
we have two distinguished groups of van Geemen lines, monodromy around $z_{\rm D}$
exchanges the two branches that we labelled~$C_\omega$ and~$C_{\omega^2}$. This
will be crucial for what follows.

We now describe the problem that we will solve in this work. The dif\/ferential equation~(\ref{deq}) admits four solutions that are linearly independent over~$\CC$. Solutions
corresponding to an integral basis of~$H_3(Y,\ZZ)$ are however well-known. Such a basis
was f\/irst obtained in~\cite{cdgp} by calculating the transformation properties of
the solutions under monodromy around the moduli space, and has played an important
role in several subsequent developments. Later, the integral variations of Hodge
structure underlying these calculations have been identif\/ied and fully classif\/ied
in this case, see~\cite{doranmorgan}, which fact provides a f\/irm basis for
our present discussion.

A similar issue af\/f\/licts the inhomogeneous equation~(\ref{iPF}). While $\tau(z)$ is
initially def\/ined in~(\ref{integral}) by the choice of chain up to an integral
three-cycle, the dif\/ferential equation determines $\tau(z)$ only up to a complex
linear combination of periods. By the same token, encircling one of the singularities
in the moduli space must return the chain up to an integral cycle so $\tau(z)$ must
transform by an integral period. Note that the availability of the above-mentioned
integral lift of monodromy is crucial for asking a sensible question here.

We will f\/ix the choice of chain equivalent to imposing specif\/ic
boundary conditions at the open string discriminant $z_{\rm D}$. As $z$ approaches
$z_{\rm D}$, pairs of solutions of (\ref{pars}) will approach each other. In $Y$, this
means that the associated van Geemen lines will coincide, and we may choose to def\/ine~$\tau(z)$
using the corresponding vanishing three-chain. Analytically, this means that~$\tau(z)$ as
well as its f\/irst derivative $\tau'(z)$ vanish at $z=z_{\rm D}$. (The f\/irst derivative
with respect to $z$ gives the integral of the f\/irst order deformation of the holomorphic
3-form, and must vanish together with the integral of $\Omega$ itself.) To see that
this condition is compatible with the behaviour of the inhomogeneity in (\ref{iPF}), we
note that in a local coordinate $y=z-z_{\rm D}$, the Picard--Fuchs equation will take
the leading order form
\begin{gather*}
\call\tau \sim \frac{d^4}{dy^4}\tau \propto y^{-5/2},
\end{gather*}
so the solution $\tau\sim y^{3/2}$ as well as $\tau'\sim y^{1/2}$ will vanish as
$y\to 0$. In fact, since as noted above, monodromy around $y=0$ exchanges~$C_\omega$
with $C_{\omega^2}$, our $\tau$ is af\/fected precisely by a change of sign. Thus,
the condition that $\tau/y^{3/2}$ be analytic at $y=0$ completely f\/ixes the
solution of the inhomogeneous equation~(\ref{iPF}).

Our main results are $(i)$ the verif\/ication that under this boundary condition at~$z_{\rm D}$,
the monodromies of $\tau$ around the other singular points are indeed integral, and $(ii)$~the calculation of the leading asymptotic behaviour at the large volume point. General
theory, see~\cite{ggk}, restricts the limiting values of normal functions under
degenerations such as that at~$z_{\rm LV}$, and we will f\/ind that our results are
compatible with these restrictions.

Calculations similar to the one described here were carried out for a dif\/ferent
class of algebraic cycles on the mirror quintic in~\cite{opening}.

\section{Expansions}

In this section, we collect the expansions of the solutions of the homogeneous
and inhomogeneous Picard--Fuchs equation around the various distinguished points
in moduli space. We begin at large volume.

Around $z=z_{\rm LV}$, a basis of solutions of equation~(\ref{deq}) can be generated
using Frobenius method from the hypergeometric series
\begin{gather*}
\varpi(z;H) = \sum_{n=0}^\infty \frac{\Gamma(5(n+H)+1)}{\Gamma(n+H+1)^5} z^{n+H}.
\end{gather*}
Indeed, since $\call\varpi(z;H)\propto H^4$, and $[\partial_H,\call]=0$, a basis of
solutions is given by
\begin{gather*}
\varphi_k(z) = \frac{1}{(2\pi i)^k} (\partial_H)^k\mid_{H=0}\varpi(z;H) , \qquad k=0,1,2,3,
\end{gather*}
with leading behaviour
\begin{alignat*}{3}
& \varphi_0(z)  = f_0  ,\qquad && f_0 = 1+120 z+113400 z^2 + \cdots , & \\
& (2\pi i)\varphi_1(z)  = f_0 \log z + f_1  ,\qquad && f_1 = 770 z +810225z^2 +\cdots, &  \\
& (2\pi i)^2\varphi_2(z)  =f_0 \log^2 z +2 f_1 \log z + f_2  ,\qquad && f_2 =
  \tfrac 25 \cdot 2875 z+ \tfrac{4208175}{2} z^2 + \cdots, &  \\
& (2\pi i)^3\varphi_3(z)  = f_0 \log^3 z+ 3 f_1 \log^2 z+3 f_2 \log z+ f_3  ,\quad
 && f_3 =  -\tfrac {12}5\cdot 2875 z- \tfrac{9895125}{2} z^2 + \cdots. &
\end{alignat*}
This basis is close to being integral, but there are some important rational as well
as irrational modif\/ications. By results of~\cite{cdgp}, a (projectively) integral
basis is given by the period vector $\Pi= (\varpi_0,\varpi_1,\varpi_2,\varpi_3)^T$
with
\begin{gather}
\varpi_0 (z)  = \varphi_0(z), \qquad
\varpi_1(z)  = \varphi_1(z),\qquad
\varpi_2(z)  = \frac 52\varphi_2(z) - \frac 52 \varphi_1(z) - \frac{25}{12}\varphi_0(z),\nonumber\\
\varpi_3(z)  = -\frac 56 \varphi_3(z) -\frac{25}{12}\varphi_1(z) + 200\frac{\zeta(3)}{(2\pi i)^3}
\varphi_0(z).\label{lvbasis}
\end{gather}
Under $z\to e^{2\pi i}z$, we have $\varphi_k\to \sum\limits_{j=0}^k \binom{k}{j} \varphi_j$ and thereby, large volume monodromy is represented on the
period vector
\begin{gather*}
\Pi \to M_{\rm LV} \Pi
\end{gather*}
by the matrix
\begin{gather*}
M_{\rm LV} = \begin{pmatrix} 1 & 0 & 0 & 0 \\
1 & 1 & 0 & 0 \\
0 & 5 & 1 & 0 \\
-5 & 5 & 1 & 1
\end{pmatrix}.
\end{gather*}
We draw attention to the one irrational constant in equation~(\ref{lvbasis}),
\begin{gather}
\label{zeta3}
200 \frac{\zeta(3)}{(2\pi i)^3} \approx i \cdot 0.9692044901\dots,
\end{gather}
which decouples from monodromy considerations at large volume, but can be determined
by a~close examination of the conifold locus (see~\cite{cdgp}).

In the local variable $w=1-5^5z$, which vanishes at $z=z_{\rm C}$ we may work
out a basis of solutions of equation~(\ref{deq}) to be given by
\begin{gather}
\psi_0  =   1+\tfrac{2}{625} w^3+\tfrac{97}{18750} w^4 +\cdots,\nonumber \\
\psi_1  =   \tfrac{\sqrt{5}}{2\pi i}\cdot\left(-w-\tfrac{7}{10} w^2
-\tfrac{41}{75} w^3-\tfrac{1133}{2500} w^4-\cdots \right),\nonumber \\
\psi_2  =
\tfrac{1}{2\pi i} \psi_1 \log w +
\tfrac{\sqrt{5}}{\pi^2}\cdot\left(-\tfrac{23}{1440} w^3-\tfrac{6397}{240000} w^4-\cdots\right),\nonumber\\
\psi_3  =   w^2+\tfrac{37}{30} w^3 +\tfrac{2309}{1800} w^4-\cdots,\label{cbasis}
\end{gather}
and assemble it into the conifold period vector $\Psi=(\psi_0,\psi_1,\psi_2,\psi_3)$.
We emphasize that this is not an integral basis of periods. In fact, it is not known
analytically how the integral basis~(\ref{lvbasis}) is related to the basis~(\ref{cbasis}). It is known however, and this explains the choice of integration
constants, that $\varpi_3=\psi_1$, and that $\varpi_0-\psi_2$, $\varpi_1$, $\varpi_2$ are
linear combinations of $\psi_0$, $\psi_1$, $\psi_3$. In other words, the conifold monodromy
$w\to e^{2\pi i} w$ is represented on the large volume period vector as
\begin{gather*}
M_{\rm C} = \begin{pmatrix} 1 & 0 & 0 & 1\\
0 & 1 & 0 & 0 \\
0 & 0 & 1 & 0 \\
0 & 0 & 0 & 1
\end{pmatrix}.
\end{gather*}
These results were originally obtained in~\cite{cdgp} by exploiting the analytic
continuation to the Gepner point, which can be done in closed form. Since to reach
the open string discriminant, we will have to resort to numerical continuation anyway,
we will skip this part of the story. We only record that on the periods, monodromy
around the Gepner point is equivalent to the composition of large volume and conifold
monodromy. It is therefore represented by the order 5 matrix
\begin{gather*}
M_{\rm G} = M_{\rm LV}\cdot M_{\rm C} =
\begin{pmatrix} 1 & 0 & 0 & 1\\
1 & 1 & 0 & 1 \\
0 & 5 & 1 & 0 \\
-5 & -5 & -1 & -4
\end{pmatrix}  , \qquad (M_{\rm G})^5 = {\bf 1}.
\end{gather*}

Turning then to the inhomogeneous equation (\ref{iPF}), we f\/ind the power series
expansion of a~solution around large volume
\begin{gather}
\label{tauLV}
\tau_{\rm LV}(z) = \tfrac{\sqrt{-3}}{\pi^2}\cdot
\left( 140000 z+\tfrac{11521900000}{3} z^2+\tfrac{5187112292000000}{27} z^3+\cdots\right)
\end{gather}
and conifold
\begin{gather}
\label{tauC}
\tau_{\rm C}(w) = \tfrac{\sqrt{5}}{\pi^2}\cdot
\left(\tfrac{88}{15625}w^3+\tfrac{4282}{390625}w^4 + \cdots\right).
\end{gather}
Finally, around the open string discriminant, $z=z_{\rm D}$ we use the variable
$y=1-2^7\cdot 5^5\cdot 3^{-1}z$. As we have explained above, we f\/ix boundary conditions
such that $\tau/y^{3/2}$ is analytic as $y\to 0$. This solution has the power series
expansion
\begin{gather}
\label{tauD}
  \tau(y) = \tfrac{\sqrt{-3}}{\pi^2}\cdot\left(\tfrac{16}{5}y^{3/2}+\tfrac{8352}{3125}
y^{5/2}+\tfrac{6156432}{2734375} y^{7/2} + \cdots\right).
\end{gather}

\section{Continuations}

The power series (\ref{tauLV}), (\ref{tauC}), and (\ref{tauD}) all represent solutions of
the inhomogeneous Picard--Fuchs equation (\ref{iPF}). These solutions must be equal modulo
a solution
of the homogeneous equation, i.e., a period. As we have explained, the solution of our
interest is determined by the boundary condition at the open string discriminant, i.e.,
$\tau\sim y^{3/2}$, and we wish to relate it to the other two expansions. (The continuation
to the Gepner point can then be inferred from this data.)

Our f\/irst task then is to express $\tau$ in terms of $\tau_{\rm LV}$ and $\Pi=
(\varpi_0,\varpi_1,\varpi_2,\varpi_3)$. For this aim, we note that the power series
(\ref{tauLV}) and
(\ref{tauD}) have a radius of convergence equal to $|z_{\rm D}|$, the latter because of
the singularity of the dif\/ferential operator~(\ref{oPF}),
and the former because of the apparent singularity in the inhomogeneity~(\ref{iPF})
 (cf.\ Fig.~\ref{space}).
Both expansions, as well as~(\ref{lvbasis}), converge well at the midpoint, $z=z_{\rm D}/2$,
which is therefore a convenient point to compare. By evaluating numerically all those
functions and their derivatives, we f\/ind that
\begin{gather}
\label{wefind}
\tau = \tau_{\rm LV} + v_{{\rm LV}}\cdot{\Pi},
\end{gather}
where
\begin{gather}
\label{vLV}
v_{{\rm LV}} = (a,-10,0,2),
\end{gather}
with
\begin{gather}
\label{zeta}
a \approx i \cdot 13.36856103560663627\dots.
\end{gather}

It is not quite that straightforward to compare $\tau_{\rm C}$ with $\tau$. While equation~(\ref{tauC}) converges up to the open string discriminant, the expansion~(\ref{tauD})
still has radius of convergence $|z_{\rm D}|$, which is much less than the distance
$|z_{\rm C}-z_{\rm D}|$. A~comparison would of course be possible in the intersection
of the two disks, but~(\ref{tauC}) does not converge well in this region. Instead,
we resort to solving the dif\/ferential equation numerically, with initial conditions
around $z_{\rm D}$ given by~$\tau$, and compare with~$\tau_{\rm C}$ close to~$z_{\rm C}$. We f\/ind
\begin{gather*}
\tau = \tau_{\rm C} + \tilde v \cdot \Psi,
\end{gather*}
where $\tilde v= (3.3655477\dots, i \cdot 2.9120714\dots,8,0.18666620\dots)$ is a certain
vector with one integral and three apparently irrational entries. Since $\Psi$ is
not fully integral, it appears more signif\/icant, and more convenient to compute
monodromies, to give the relation to the integral large volume basis, namely,
\begin{gather*}
v_{\rm C} = \tilde v \Psi\cdot \Pi^{-1} = (8,c_1,c_2,c_3),
\end{gather*}
where
$c_1\approx 9.197317\dots- i\cdot 13.598567\dots$,
$c_2\approx 3.6789269\dots$,
$c_3\approx - i \cdot 2.3915634\dots$.

With all this data in hand, it is now easy to write down the transformation properties
of our distinguished solution $\tau$ under monodromy about the various singularities.
Using~(\ref{wefind}) and the fact that $v_{\rm LV}$ is single valued around large volume,
we obtain
\begin{gather*}
\tau\to\tau + m_{\rm LV}\cdot \Pi,
\end{gather*}
with
\begin{gather*}
m_{\rm LV} = v_{\rm LV}\cdot M_{\rm LV} - v_{\rm LV} = (-20,-10,2,0).
\end{gather*}
About the conifold, we exploit invariance of $\tau_{\rm C}$ to f\/ind that
\begin{gather*}
\tau\to\tau+ m_{\rm C}\cdot \Pi,
\end{gather*}
with
\begin{gather*}
m_{\rm C} = v_{\rm C}\cdot M_{\rm C}- v_{\rm C} = (0,0,0,8).
\end{gather*}
And, of course, monodromy around the open string discriminant is, by construction
\begin{gather*}
\tau \to -\tau.
\end{gather*}
Putting this together, we may infer also the monodromy around the Gepner point
\begin{gather*}
\tau\to -\tau+ m_{\rm G}\cdot \Pi,
\end{gather*}
with
\begin{gather*}
m_{\rm G} = m_{\rm C} + m_{\rm LV}\cdot  M_{\rm C} = (-20,-10,-2,-12).
\end{gather*}
One may also check that the order of the Gepner monodromy has been extended
from~5 to~$10$, rather as in~\cite{opening}.

\section{Discussion}

The basic result of the previous section is the integrality of the vectors $m_{\rm LV}$,
$m_{\rm C}$, $m_{\rm G}$. Under monodromy around the moduli space, the truncated normal
function comes back to itself (or minus itself, if one encloses the open string discriminant
or Gepner point, which are the branch points of the inhomogeneity equation~(\ref{iPF})),
up to an integral period. It remains to discuss the signif\/icance of the one irrational
entrie in $v_{\rm LV}$, given in equation~(\ref{zeta}).

The general theory of limiting values of normal functions under degeneration of the
underlying variation of Hodge structure has been recently explained by Grif\/f\/iths, Green
and Kerr~\cite{ggk}. This work provides, f\/irst of all, the identif\/ication of the
Abelian group ``f\/illing in'' for the intermediate Jacobian at the singular f\/iber
of a degenerating Hodge structure. These ``N\'eron models'' then allow the discussion of
the limiting values of normal functions. And for a normal function coming from geometry
(i.e., an algebraic cycle), this limiting value is interpreted as the Abel--Jacobi
map on a certain motivic cohomology of the singular f\/iber. An important
point is that the theory of~\cite{ggk} deals with the phenomena in the strict
degeneration limit, whereas we have been concerned here with the analytic
expansion of the normal function around such a limit.

In somewhat simplif\/ied language, and specialized to the case of our interest
(co-dimension 2 cycles on a one-parameter family of Calabi--Yau threefolds, in a
maximal unipotent degeneration) the main result of~\cite{ggk} states that, in the
limit $z\to 0$, any normal function arising from an algebraic cycle must asymptote
to the period of the one monodromy-invariant three-cycle. Remembering that this is
just the fundamental period $\varpi_0(z)$, the result we found in equation~(\ref{vLV})
is in precise agreement with that condition. (All integral periods would disappear
under the map to the intermediate Jacobian.) Intuitively, one may understand
the limiting condition from the requirement that the monodromy be integral. In this
respect, the coef\/f\/icient $a$ is very similar to the constant of equation~(\ref{zeta3})
appearing in the expansion of the ordinary periods. It should also be pointed out
that the condition holds as stated under the assumption that the algebraic
cycle itself is invariant under the large volume monodromy. Otherwise,
one might pass to an appropriate cover of the moduli space, with results
similar to those in~\cite{opening}.

It is also explained in~\cite{ggk} that the limiting value of the normal
function (i.e., the coef\/f\/icient of the fundamental period) has an arithmetic
signif\/icance. Namely, our $a$ is the image of the limiting cycle under a certain Bloch
regulator map. As such, it should be expressible in terms of the Bloch--Wigner
function
\begin{gather*}
D(z) = {\rm Im}\bigl( {\rm Li}_2(z)\bigr) + {\rm arg}(1-z)\log|z|.
\end{gather*}
Indeed, guided by the example of~\cite{ggk}, we f\/ind (numeratically), that
\begin{gather}
\label{verify}
a = \frac{195}{\pi^2}D\big(e^{2\pi i/3}\big).
\end{gather}
An alternative expression follows from the relation between special values of the
Bloch--Wigner function and special values of L-functions of algebraic number f\/ields,
see~\cite{zagier}. In the case at hand, we recall that the def\/inition of our
van Geemen lines involved from the very beginning that number f\/ield $\QQ(\sqrt{-3})$.
The corresponding Dirichlet L-function is
\begin{gather}
\label{alternate}
L(s) = \sum_{k=1}^\infty \left(\frac{k}{3}\right) \frac{1}{k^s}
\end{gather}
and using formulas of~\cite{zagier}, we obtain
\begin{gather}
\label{obtain}
a = \frac{195\sqrt{-3}}{2\pi^2} L(2).
\end{gather}
The main reason for giving this alternate formula is that it allows
making contact with the results of~\cite{arithmetics}
concerning the arithmetic properties of the actual expansion of the normal function
around $z=0$ (as opposed to just its limiting value). Indeed, it was observed that
to exhibit the underlying integrality of that expansion under the mirror map, one
has to twist the standard Ooguri--Vafa di-logarithm multi-cover formula precisely as
in equation~(\ref{alternate}). This was dubbed the D-logarithm,
\begin{gather*}
{\rm Li}_2^{(\chi)} (z) = \sum_{k=1}^\infty \left(\frac{k}{3}\right) \frac{z^k}{k^2}
\end{gather*}
with the obvious coincidence of special values ${\rm Li}_2^{(\chi)}(1)= L(2)$.
We do not know whether these relations generalize.

To close, we point out two obvious questions that deserve immediate attention.
First of all, one should verify the formula~(\ref{verify}) or~(\ref{obtain}) from
direct geometric considerations, closely related to the example of~\cite{ggk}.
Secondly, one should aim to understand the origin of these special values from
the A-model or space-time physics perspective. There are two rather plausible
explanations: From the point of view of a Lagrangian D-brane that is mirror to
the algebraic cycle we have been considering, the constant $a$ should be
related to a certain perturbative invariant in the Chern--Simons theory
living on that D-brane. These invariants are known to take arithmetric values,
see e.g.,~\cite{sergei}. Lastly, from the point of view of the topological
sigma-model, $a$ should be related to a perturbative $\alpha'$ correction to
the bosonic potential of the K\"ahler moduli f\/ields induced by the presence of
the D-brane. This would be similar to the way that the constant
$\chi(X)\zeta(3)/\pi^3$ in~(\ref{zeta3}) arises from a four-loop correction to
the sigma-model metric on the background Calabi--Yau $X$, see~\cite{cdgp}.
To the best of our knowledge, the corresponding open string calculation has
not yet been done.

\subsection*{Acknowledgements}

We would like to thank Matt Kerr for asking the question addressed in this work,
and Josh Lapan for stimulating discussions. J.W.\ wishes to thank the Simons
Center for Geometry and Physics, where this paper was written up.
This work was supported in part by
the Canada Research Chair program and an NSERC discovery grant.

\pdfbookmark[1]{References}{ref}
\LastPageEnding

\end{document}